\documentclass[aps,pra,
twocolumn,
superscriptaddress,showpacs]{revtex4-2} 
\usepackage{graphicx}
\usepackage{gensymb}
\begin{document}
\title{How a simple pendulum inside a running elevator oscillates    
%Using a running elevator to effectively realize a variable gravitational acceleration 
}

\author{Mingyuan Shi}

\address{Fudan International School, Shanghai  200433, China} 

\author{Yu Shi}

\address{%USTC Shanghai Institute for Advanced Studies,  Shanghai 201315, China %\\   
  Department of Physics, Fudan University, Shanghai 200438, China}

%\begin{history}
%\received{21 April 2014}
%\accepted{21 April 2014}
%\published{13 June 2014}
%\end{history}

\begin{abstract}
 We propose to effectively realize a time-dependent gravitational acceleration by using a running elevator, so that a simple pendulum inside it effectively becomes one with a time-dependent gravitational acceleration. We did such an experiment using a realistic elevator, and analyzed the data. The acceleration of an elevator is much smaller than the gravitational acceleration, and is time-dependent only when the elevator starts and stops. However, we have managed to establish the effect on the oscillation of the pendulum. The effect becomes pronounced if the simple pendulum is put in a container vertically accelerating, and the acceleration is time-dependent, while its magnitude is comparable with that of the gravitational acceleration.  
 
\end{abstract}
 
%\usepackage{CJK}
%\usepackage{amsmath,amsfonts,amssymb,mathrsfs,bm}
%\usepackage{graphicx,graphics,color,epsfig}
%\begin{document}

\maketitle
  
\section{Introduction}
 
How does a simple pendulum oscillate if the gravitational acceleration varies with time? On the time and space scales of the oscillation of the pendulum, the gravitational acceleration is usually regarded as a constant. Although the gravitational potential of the earth, and thus the gravitational acceleration, do depend on the distance from the center of the mass of the earth, the time and space scales of the variation are far beyond those of the oscillation of the pendulum, hence irrelevant to the question above. For a pendulum hung to a static point, the variation of its weight  is negligible. So a pendulum with a variable gravitational acceleration seems hard to realize.  

In this paper, however, we propose to effectively realize a variable gravitational acceleration by putting the simple pendulum inside a running elevator, whose acceleration varies during its movement. In the reference frame of the elevator, which itself accelerates with variable acceleration, the pendulum is   subject not only to its weight but also to a fictitious inertial force due to the acceleration of the elevator with respect to the ground. The sum of these two forces leads to an effectively variable gravitational acceleration. We have performed an experiment on a simple pendulum inside a running elevator, and the data are analyzed and compared with the theory.

In Section 2, we present the theoretical reasoning and derivation regarding the realization of a variable gravitational constant. In Section 3, we describe the motion of the elevator. In Section 4, we describe our experiment. In Section 5, we describe the experimental data. In Section 6, we analyze the oscillation of the simple pendulum in our experiment. Conclusion and summary are made in Section 7. 

\section{Theory}

\begin{figure}[b]
\centerline{\includegraphics[width=8cm]{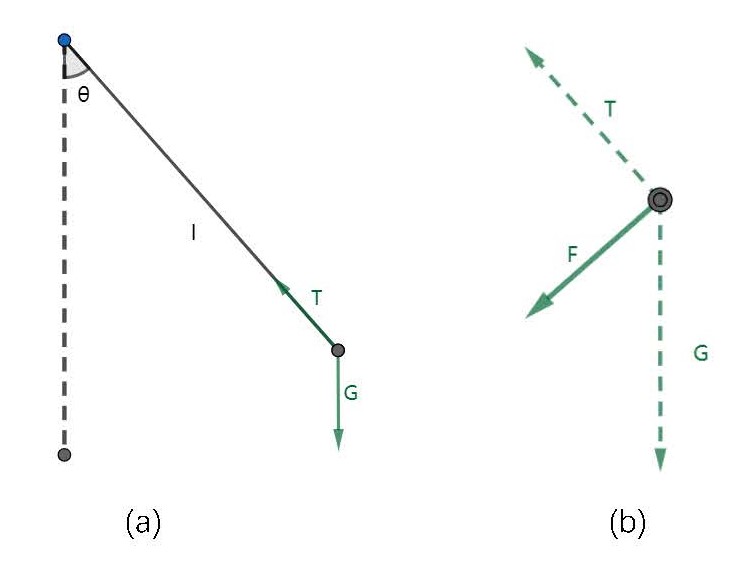}} 
\caption{ (a) A simple pendulum. (b) The combined force on the ball.     \label{fig1}}
\end{figure}

First, consider a usual simple pendulum on the ground  (Fig.~\ref{fig1}).  Suppose it consists of a small ball with mass $m$, and a string of length $l$ with negligible mass, and that the string deviates from the vertical direction for angle $\theta (t)$ at time $t$. The ball is subject to its weight  $G \equiv mg$, where $g$ is the gravitational constant, and a force of tension. The two forces are combined into the total force
$ F(t)= mg\sin\theta(t)$,   as shown in Fig.~\ref{fig1}.  
  
For the tangent motion, Newton's  Second Law gives
$$mg\sin\theta (t) = -ml \frac{d^2\theta (t)}{dt^2}. $$    
When $\theta (t)$  is small, $\sin\theta (t) \approx \theta(t)$, we have 
\begin{equation}
\frac{d^2 \theta (t)}{dt^2} =-\frac{g}{l} \theta(t). \label{eq1}
\end{equation} 
The solution tells that the angle $\theta(t)$ is periodic in time,	  
\begin{equation}
\theta(t)=\theta_m  \sin ( \omega t+\phi),	     
\label{eq2} 
\end{equation}
with $\omega =\sqrt{g/l}$.     The periodicity is $T=2\pi / \omega=2\pi \sqrt{l/g}$. Mathematically, when $g$  depends on time and becomes $g(t)$, the equation of motion becomes  
\begin{equation}
\frac{d^2 \theta (t)}{dt^2} =-\frac{g(t)}{l} \theta(t),
\label{eq3} 
\end{equation}
where $g(t)$ is now the variable gravitational constant. In general, one cannot simply use the  formula (2), with $g$ replaced as time-dependent. Instead, one needs to solve the equation of motion  (\ref{eq3}) .

But how to realize the time-dependent $g(t)$?  Here we propose to use a running elevator. Let us represent the acceleration of the elevator with respect to the ground as $-\mathbf{a}(t)$. Then in the reference frame of the elevator, apart from the tension, the force acting on the pendulum is the sum of the weight $m\mathbf{g}$ and the inertial force $m\mathbf{a}(t)$~\cite{kittle}, and can be written as $m\mathbf{g}(t)$, the magnitude of  which  is 
\begin{equation}
g(t)=g+a(t),   
\label{eq4} 
\end{equation}
which replaces $g$. Therefore $g(t)$ is the effective gravitational acceleration felt by the pendulum in a running elevator. 

This result can be obtained in a second way. Inside an elevator, the simple pendulum, besides its oscillation with respect to the elevator, also moves together with the elevator. Therefore, $-m\mathbf{a}(t)$  should be deducted from the weight  $m\mathbf{g}$, and the remaining part, as the cause of the oscillation, replaces the weight  $m\mathbf{g}$   in governing the oscillation. The remaining part is $m\mathbf{g}-[- \mathbf{a}(t)]= m[\mathbf{g} +\mathbf{a}(t)]$.  So $\mathbf{g} +\mathbf{a}(t)$ replaces $\mathbf{g}$ in the equation of motion of $\theta(t)$.  

Furthermore, this result can also be obtained in a third way, using the reference frame of the ground. In addition to the oscillation, the pendulum also moves together with the elevator, with the acceleration $-\mathbf{a}(t)$, whose component along the tangent direction of the oscillation is $-a(t)sin\theta(t)$. Therefore, the equation of motion along the tangent direction is 
\begin{equation}
mg\sin \theta(t) =  m[-l \frac{d^2 \theta (t)}{dt^2} -m a(t) \sin\theta(t)],
\label{eq5}   
\end{equation}
which gives rise to
\begin{equation}
  \frac{d^2 \theta (t)}{dt^2} =-\frac{g+a(t)}{l} \theta(t).   
\label{eq6} 
\end{equation}
We can regard the downward direction as positive, hence $g$  is positive. If the acceleration $-\mathbf{a}(t)$  of the elevator is downwards, $-a(t)$ is positive, hence $a(t)$ is negative. This is the case of weight loss. The case of overweight is opposite.

For a running elevator, $a(t)$ is not known beforehand. In our experiment described below, both $a(t)$ and  $\theta(t)$  are measured, and a comparison is made between the measured values of  $\theta(t)$  and the sine function for the constant g, and they are found to be different.

Theoretically, with $g(t)=g+a(t)$  input to the equation of motion of $\theta(t)$, Eq. (\ref{eq3}), the solution  $\theta(t)$  describes the motion of the simple pendulum with respect to the elevator. To numerically solve Eq. (\ref{eq6}) directly using the measured values of $g+a(t)$, the data points need to be dense enough, which is not easy, especially in measuring $\theta(t)$. Instead, we find an analytical function fitting the data of $1+(a(t))/g$, which is then substituted to the equation of motion of  $\theta(t)$  and the solution of  $\theta(t)$  is found and compared with the experimental result. In this way, the theory is tested. 

\section{Motion of The Elevator}

Let us first discuss the motion of the elevator, with variable acceleration with respect to the ground. 

Suppose the elevator moves downwards. For a downwards movement,  at the beginning, the acceleration of the elevator $-\mathbf{a}(t)$ is downwards and its magnitude increases, the speed increases with increasing rate.  

Then the magnitude of acceleration of elevator decreases to zero, but the speed continues to increase to reach the highest speed downwards, because the acceleration remains downwards although the growth rate of the speed slows down. 
Afterwards, the elevator stops speeding up and keeps this speed, so it performs uniform linear motion, $a(t)=0$.

In order to stop at a certain floor, the elevator has to slow down, so the direction of acceleration is opposite to the direction of the velocity. Now the direction of the velocity is still downwards, so the direction of the acceleration is upwards and the magnitude first increases and then decreases.  

If the elevator moves upwards, the process is similar to the case of moving downwards, but in each instant, the directions of the accelerations and velocities are both opposite to the downwards case.

When a simple pendulum is inside the elevator, as shown above, the acceleration $-\mathbf{a}(t)$ of the running elevator affects the oscillation of the pendulum during the motion of the elevator through the effective gravitational acceleration $\mathbf{g}(t)=\mathbf{g}+\mathbf{a}(t)$ felt by the pendulum, replacing $\mathbf{g}$ for a simple pendulum on the ground or in a static elevator.   
 
 Actually, with respect to the elevator, the apparent weight of any object of mass M becomes $M[\mathbf{g}+\mathbf{a}(t)]$, where $Ma(t)$ can be understood as an inertial force in the non-inertial frame.  
 
Due to this fact,    $\mathbf{a}(t)$  can be obtained by using the apparent weight $\mathbf{W}$ measured in the experiment through the equation $\mathbf{a}(t)=\mathbf{W}/m-\mathbf{g}$.  In our experiment, for convenience, the simple pendulum and the timer are put on the balance, with the total weight $\mathbf{W}$, the reading gives $W/g$. When the elevator is static, it is just $m$. In any case, the ratio $W/mg$  is exactly $1+ a(t)/g$.

Previously we have measured the  acceleration of an elevator according to the apparent weight of an object inside it~\cite{shi}. 

\section{Experiment}

\begin{figure}[b]
\centerline{\includegraphics[width=7cm]{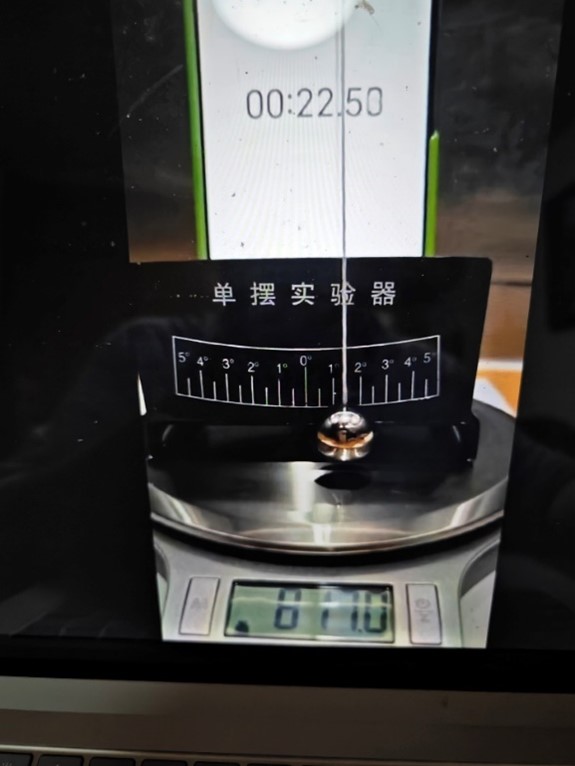}} 
\caption{ 
A screenshot of the video of the process of the experiment. 
\label{fig2}}
\end{figure}

In addition to the elevator, our experimental devices and materials include: a simple pendulum with 	a scale with the maxima of $5\degree$ on left and right sides. and with a minimum graduation of $0.5\degree$ (Fig.~\ref{fig2}); 	a steel ball with a radius of $0.95$  cm; 	a string with length $37.4$ cm, a balance with a measurable range between $1$ gram to $5$  kilogram, and accuracy of $0.1$ gram; a timer with accuracy of $0.01$ second, or a smartphone with such a timer; a camera equipment, or another smartphone with a camera.  Hence in our experiment,  $l= 37.4+0.95=38.35$ cm, which is from the top of the string to the center of the ball. 

We do an experiment on a simple pendulum inside the  elevator, and study the relation between the oscillation of the simple pendulum and the acceleration $-\mathbf{a}(t)$ of the elevator. As explained above, the effective gravitation acceleration for the simple pendulum is now $\mathbf{g}+\mathbf{a}(t)$.   
 
We put the pendulum and the timer on the balance inside an elevator. The reading of the  total  apparent weight gives the information on $g+a(t)$.   
Turn on the camera to record the video of the whole process of the experiment.   The timer is going on, so the timing of the motion of the simple pendulum can be recorded accurately.  Release the ball of the simple pendulum, with the string straightened, at a particular angle between the string and the vertical direction.  Then press the button of the elevator, which moves until it stops at the certain floor.  Repeat the operation if necessary. 

In this way, the data collection is completed.  In the experiment analyzed in this paper, the elevator moves between Level 6 and Level 1. Throughout the observation of the experiment, the simple pendulum oscillates. 

Fig.~\ref{fig2}  is a screenshot of the video which illustrates a device of a simple pendulum, a timing timer, and a balance in the process of the experiment.

\section{Experimental Data}
 
From the video record, we extract the data for a duration of time in which the elevator moves from Level 6 to  Level 1.  We obtain the apparent weight $m[g+a(t)]$ and the oscillating angle  $\theta(t)$  as functions of time t. Hence the ratio between apparent weight and the real weight, $1+a(t)/g$, is obtained.  In particular, when the elevator is static, the weight is just $mg$, so the mass is found to be  $816 gram$. 

\begin{figure}[b]
\centerline{\includegraphics[width=8cm]{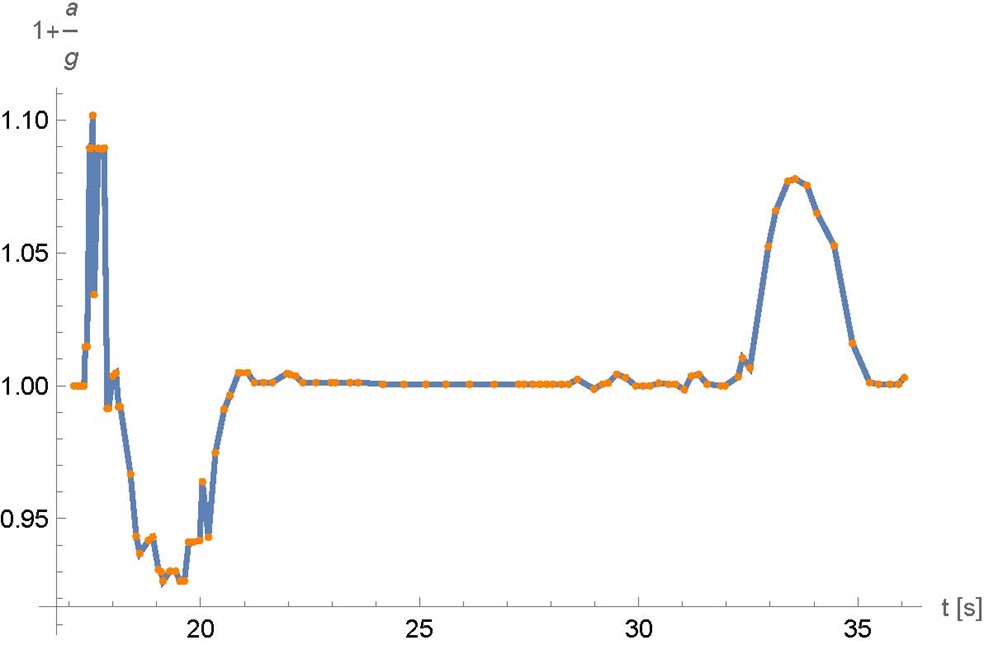}} 
\caption{ The ratio between apparent weight and the real weight, $1+a(t)/g$, as a function of time $t$.   
\label{fig3}}
\end{figure}

\begin{figure}[b]
\centerline{\includegraphics[width=8cm]{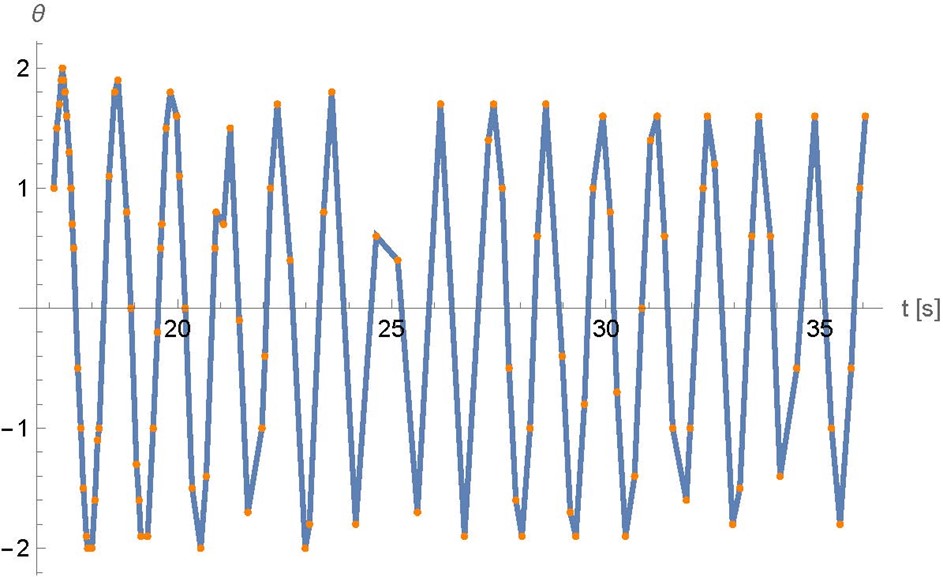}} 
\caption{ The oscillating angle  $\theta(t)$  as a function of time $t$.    
\label{fig4}}
\end{figure}

$\theta(t)$  and 1+a(t)/g are plotted respectively as functions of time t, by using the software {\it Mathematica}, as shown in Fig.~\ref{fig3} and Fig.~\ref{fig4}, respectively.

According to Fig.~\ref{fig3}, let us discuss the actual motion of this elevator.   Note that the data are from an actually running elevator, which is not ideal. There is a very brief time duration when  $a(t)$  is positive on the contrary, with seven data point, which means the acceleration of the elevator $-a(t)$  is upwards on the contrary. At about $18s$, the elevator starts to move downwards, and  $a(t)$  become negative, as expected for the starting period of downwards motion.  

From about $18s$ to $19s$, $1+a(t)/g $ declines from $1.00$ to about $0.92$, indicating  that  the elevator accelerates downwards with increasing acceleration $-a(t)$. In a following short duration of about $5$ seconds, the magnitude of acceleration reaches the largest value, which is about $0.08g$. 
Then from about $19.5s$ to $22s$, $1+a(t)/g$ increases back to about $1.00$, which means  $-a(t)$ decreases back to $0$. The direction of the acceleration remains downwards during this period. So the apparent weight is less than the real weight. 
Then from $22s$ up to about $32s$, $a(t)=0$, and the elevator moves uniformly. 
Afterwards, $1+a(t)/g$  increases to about $1.08$ at about $33.5s$, and then decrease towards $0$ at about $36s$. This indicates a process with acceleration $-a(t)$   upwards, $-a(t)<0$. The magnitude of the acceleration $-a(t)$ increases from $32s$ to $33.5s$, reaching the maximum $0.08g$, and then decreases to $0$ at $36s$. The elevator decelerates to $0$  when it is going to arrive the Level 1.

It can be seen that the process between $32s$ and $36s$ is almost symmetric with the process between $18s$ and $22s$. For $0\leq\tau \leq 3$, the accelerations at $32s+\tau $ and $18s+\tau $ are opposite in direction but equal in magnitude. 

\section{Oscillation of The Simple Pendulum}

The actual oscillation of the pendulum is displayed in Fig.~\ref{fig4}. We now compare the actual data with the solutions to Eq.~(\ref{eq6}) and Eq.~(\ref{eq1}).
The data points  read from the video record are not particularly dense, thus Eq. (6) cannot be numerically solved by directly using the experimentally measured values of  $1+a(t)/g$ at these data points.  Instead, to find the solutions to Eq. (6), we first find an analytical expression fitting the data of $1+a(t)/g$, as shown in Fig.~\ref{fig3}.

\begin{figure}[b]
\centerline{\includegraphics[width=8cm]{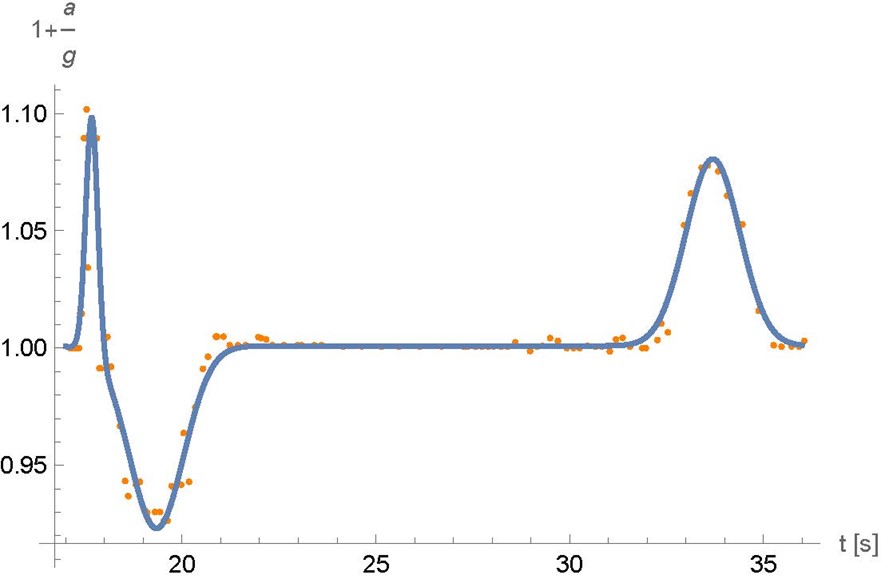}} 
\caption{  Fitting the experimental data points of  $1+a(t)/g$. The blue line is the function given in Eq.~(\ref{eq7}).   
\label{fig5}}
\end{figure}

As seen in  Fig.~\ref{fig3}, there are only a few peaks in which  $a(t)$  is nonzero, when the elevator starts or stops. The idea is to use three Gaussian functions centering at the peaks and valleys, respectively, as shown in Fig.~\ref{fig5}.

\begin{figure}[b]
\centerline{\includegraphics[width=8cm]{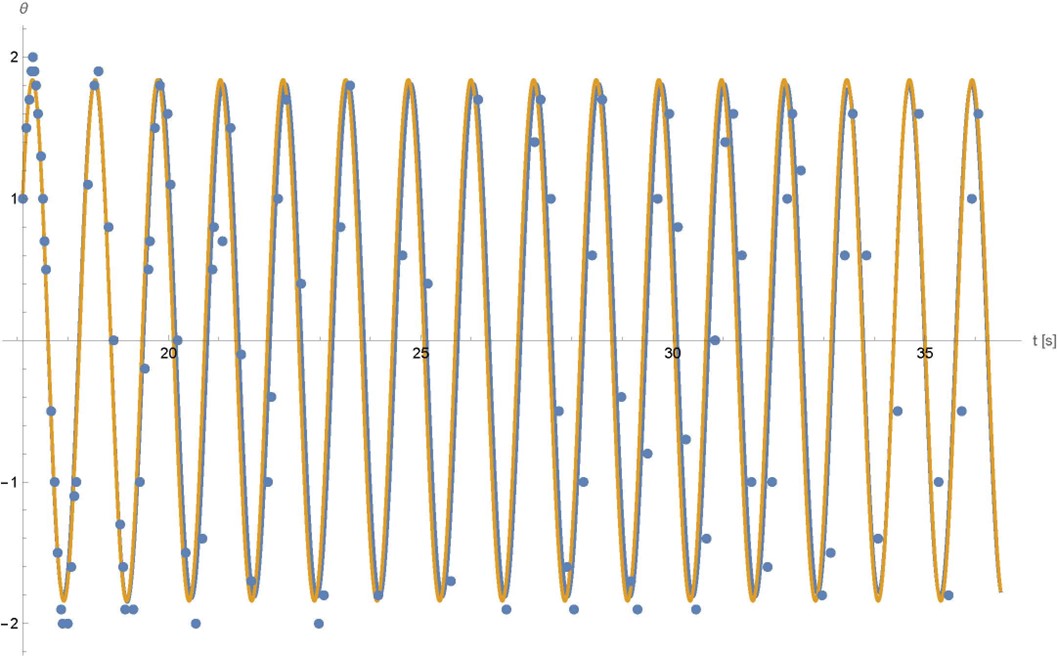}} 
\caption{
 The oscillation angle $\theta(t)$. The points are experimental data. The yellow thick curve is the solution to Eq. (\ref{eq1}), with constant $g$, the blue thin curve is the solution to Eq. (\ref{eq6}), with $g+a(t)$. 
\label{fig6}}
\end{figure}

It is found that 
\begin{equation}
\begin{array}{rl} 
1+\frac{a(t)}{g} \approx&   1.00111+0.101508 e^{20.9344(t-17.67)^2} \\ &-0.0779814e^{-1.02643(t-19.34)^2}
\\&  +0.079353e^{-1.02643(t-33.7)^2},  
\end{array}    \label{eq7}
\end{equation}
which is substituted into Eq. (\ref{eq6}). Then it is solved by using the initial condition $ \theta_0=\theta(17.11s)$, $\theta'_0=[\theta(17.24s)-\theta(17.11s)]/(17.24s-17.11s)$, where $\theta(17.11s)=1\degree$, $\theta(17.18s)=1.5\degree$,  $\theta(17.24s)=1.7\degree$,  which are read from our experimental data.  Eq. (\ref{eq1}) is also solved under the same initial condition.  Then the solutions to these two equations and the experimental data on  $\theta(t)$  are displayed together in Fig. \ref{fig6}.

It can be seen that the data points are consistent with the solution to Eq. (\ref{eq6}), and deviate from the solution to Eq. (\ref{eq1}).  It can also be seen that the two curves well coincide at the very beginning and at the very end, and there are differences in the duration of uniform motion (from $22s$ to $32s$). The curves coincide at the beginning because the effect of nonzero  $a(t)$  has not accumulated enough, they coincide again at the end because the effect of  $a(t)$  from $32s$ to $36s$ cancels that from $18s$ to $22s$, as the signs are opposite. 

In the period  of uniform motion (from $22s$ to $32s$), $a(t)=0$, so the two curves and the experimental data all correspond to a same periodic oscillation, their differences are only slight global shifts. In other words,  $\theta(t)$  at $22s$ sets the initial condition for the period  between $22s$ and $32s$, with constant $g$.

\begin{figure}[b]
\centerline{\includegraphics[width=8cm]{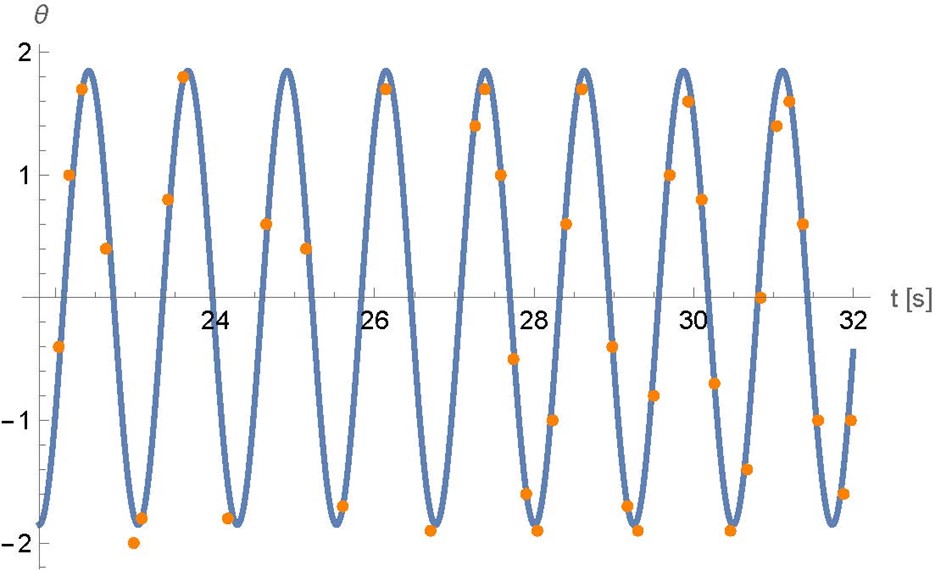}} 
\caption{ When $a(t)=0$,  $\theta(t)$  should be a sinusoidal function. The experimental data between  $22s$ to $32s$ are fitted by using a sinusoidal function.
\label{fig7}}
\end{figure}

We now fit the experimental data of  $\theta(t)$  for the period of uniform motion by using Eq. (\ref{eq2}). We find $\theta_m \approx 1.84406\degree$, $\phi \approx 1.4044$, as shown in Fig.~\ref{fig7}.

\begin{figure}[b]
\centerline{\includegraphics[width=8cm]{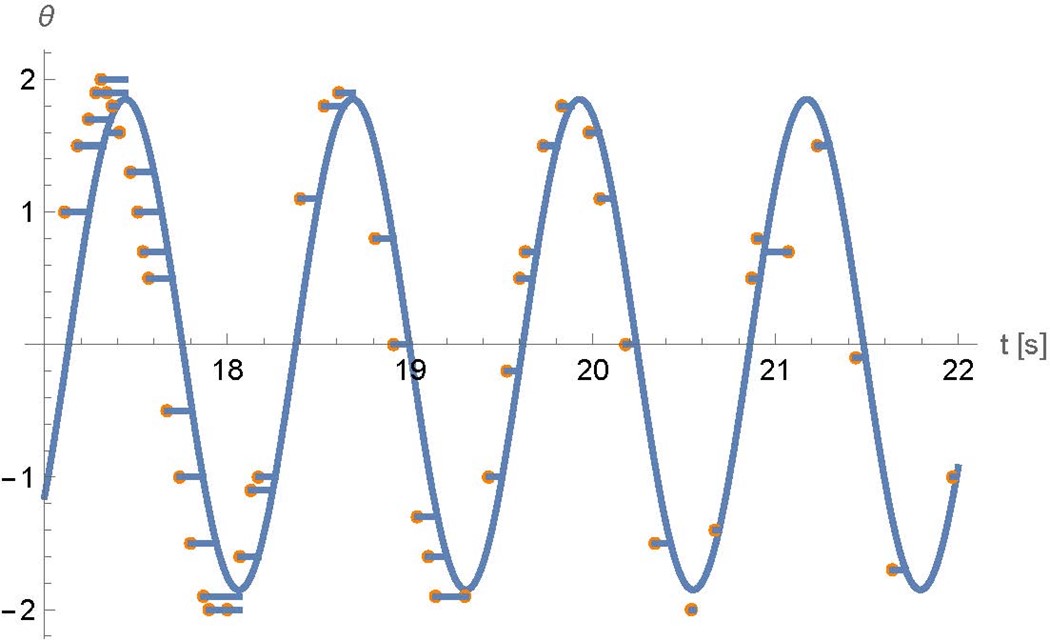}} 
\caption{ The data points of  $\theta(t)$   in the duration earlier than $22s$, in comparison with the sinusoidal function extended from that in Fig.~\ref{fig7}. 
\label{fig8}}
\end{figure}

To see the effect of nonzero  $a(t)$   at earlier times, we extend this sinusoidal function into the earlier times, as shown in Fig.~\ref{fig8}.

Now the effect of nonzero  $a(t)$  can be seen. Before $22s$, the actual oscillation is advanced compared with the sinusoidal function extended from the data between $22s$ and $32s$. The earlier, the more advanced. The closer to $22s$, the less advanced. 

From $17s$ to $18s$, the effect of nonzero  $a(t)$  is negligible because the effect has not yet been accumulated enough, so the data points, the curve of the solution with $g+a(t)$, and the curve of the solution with constant $g$ all coincide. In Fig.~\ref{fig8}, the data points from $17s$ to $18s$ are most advanced compared with  the sinusoidal function extended from that fitting the data points between $22s$ and $32s$.

In general, the solution to Eq. (\ref{eq6}) is not a simple sinusoidal function as  $a(t)$  is not a constant. However, for a short duration and for small magnitudes of $a(t)$, we can use it in an approximate way. Because the data points between $17s$ and $18s$ are most advanced with respect to the sinusoidal function extended from that fitting the data from $22s$ to $32s$, while the advance is less and less until data fit the the sinusoidal function at about $22s$, the approximate periodicity from $18s$ to $22s$ must be longer than the case of $a=0$, as the number of periodicities are the same as that of $a=0$, and they are ``stretched'' so that from $22s$ to $32s$, they fit the sinusoidal function. Therefore from $18s$ to $22s$, $a(t)<0$, as approximately $T \propto  \sqrt{l/[g+a(t)]}$  for small  $|a(t)|$.  This is consistent with the fact that indeed the acceleration of the elevator is indeed $-a(t) >0$ in this part of the movement.  

\section{Summary and Conclusion}

A simple pendulum with time-dependent gravitational acceleration was   rarely studied previously. In this paper, we propose to use the acceleration of an elevator to effectively realize the variable gravitational acceleration felt by the pendulum. This work combines theory and experiment.

In general, time-dependent gravitational acceleration $g(t)$ changes the motion of the pendulum from the simple periodic oscillation. In the approach of this paper, $g(t)=g+a(t)$,  where  $a(t)$  is minus the acceleration of the elevator. An experiment has been carried out in which  $a(t)$  and the angle  $\theta(t)$  are measured altogether. 

It turns out that the magnitude of $a(t)/g$ is very small (the largest value is only about $0.1$) and is nonzero when the elevator starts or stops. According to the equation of motion of  $\theta(t)$, there are only small differences with the case of constant $g$, the effect on  $\theta(t)$  is perturbative. Nevertheless, we have analyzed the differences although they are small.  

We have found an approximate analytical expression of $1+a(t)/g$, and then found the solution to the equation of motion with $g(t)=g+a(t)$, which is consistent with experimental data of $\theta(t)$. We have also compared the experimental data of  $\theta(t)$  with the sinusoidal function, which is the solution for the equation of motion with g. It has been shown that the experimental data of  $\theta(t)$  are consistent with the data of $1+a(t)/g$. The differences between the experimental data and the sinusoidal function are explained and analyzed. 

As the acceleration  $a(t)$  of the elevator is nonzero only when the elevator starts or stops, and its magnitude is much smaller than the gravitational acceleration g, its effect is very small. That is to say, the variability of the effective gravitational constant $g(t)$ in our experiment is limited. Nevertheless we have established the effect of an effectively variable gravitational constant. The theory and experiment are consistent.  

We have presented the general idea. Hopefully, more significant effects can be realized by putting the simple pendulum in a specifically designed container that can vertically accelerate with variable large acceleration.

%\end{CJK}
\end{document}